\documentclass[final]{article}
\usepackage{graphicx}
\usepackage{bm}
\usepackage{eucal}%
\usepackage{mathrsfs}%
\usepackage[utf8]{inputenc}

\usepackage[bookmarks=false]{hyperref}

\usepackage[caption=false]{subfig}
\graphicspath{{figures/}}
\usepackage{placeins}
\usepackage[draft]{fixme}
\usepackage{xspace}

\newcommand\meanx{\ensuremath{\left\langle x\right\rangle}\xspace}

\usepackage{mathtools}
\DeclarePairedDelimiter{\ceil}{\lceil}{\rceil}

\begin{document}

\vspace*{0.35in}

\begin{flushleft}
{\Large
\textbf\newline{Threshold games and cooperation on multiplayer graphs}
}
\newline
\\
Kaare B. Mikkelsen\textsuperscript{1,2,*},
Lars A. Bach\textsuperscript{1,3},

\bigskip
\bf{1} Interacting Minds Center, Aarhus University, DK-8000 Aarhus C, Denmark
\\
\bf{2} Department of Engineering, Aarhus University, DK-8000 Aarhus C, Denmark
\\
\bf{3} Interdisciplinary Center for Organizational Architecture (ICOA),
Aarhus University, DK-8210 Aarhus V, Denmark
\bigskip


* mikkelsen.kaare@gmail.com

\end{flushleft}

\begin{abstract}
\subsection*{Objective}
\noindent The study investigates the effect on cooperation in multiplayer games, when the population from which all individuals are drawn is structured – i.e. when a given individual is only competing with a small subset of the entire population.

\subsection*{Method}
\noindent To optimize the focus on multiplayer effects, a class of games were chosen for which the payoff depends nonlinearly on the number of cooperators – this ensures that the game cannot be represented as a sum of pair-wise interactions, and increases the likelihood of observing behaviour different from that seen in two-player games. The chosen class of games are named “threshold games”, and are defined by a threshold, $M > 0$, which describes the minimal number of cooperators in a given match required for all the participants to receive a benefit. 
The model was studied primarily through numerical simulations of large populations of individuals, each with interaction neighbourhoods described by various classes of networks. 

\subsection*{Results}
\noindent When comparing the level of cooperation in a structured population to the mean-field model, we find that most types of structure lead to a decrease in cooperation. This is both interesting and novel, simply due to the generality and breadth of relevance of the model – it is likely that any model with similar payoff structure exhibits related behaviour. 

More importantly, we find that the details of the behaviour depends to a large extent on the size of the immediate neighbourhoods of the individuals, as dictated by the network structure. In effect, the players behave as if they are part of a much smaller, fully mixed, population, which we suggest an expression for. 

\end{abstract}

\textbf{Keywords:} game theory; network; threshold; cooperation; volunteer's dilemma

\textbf{Highlights:}
\begin{itemize}
\item Observed behaviour depends on the size of each player's immediate interaction neighbourhood.
\item When the number of players is much larger than the number of required cooperators, average payoff decreases.
\item Most network structures lead to a decrease in cooperation compared to the fully mixed case.
\end{itemize}

\section{Introduction}

For the greater part of a century, the special brand of mathematics called game theory has been used to understand behaviour of social animals - including humans. Much research today still focus on mechanisms that can push short sighted self-rewarding behaviour towards behaviours less costly to conspecifics and hence more Pareto effecient for the population \cite{Hofbauer1998Evolutionary,Broom2013GameTheoretical}. While much, but far from all, cooperation and coordination among non-human species seem to coincide with kin relations \cite{CluttonBrock2009Cooperation}, human interactions seem more culturally loaded with elements of punishment, reputation and normative behavioural protocols. By providing mathematical clarity combined with recognizable narratives, game theory has played a central role in describing the nature and emergence of cooperative behaviour \cite{Szathmary1995Major} within groups and populations. In the majority of these studies, the population being studied is assumed to have no structure, meaning that all individuals may interact directly with all others, at random, and the interactions themselves are assumed to be pairwise, rather than true multi-participant dynamics. These are all useful simplifications, as already pointed out in \cite{Motro1991Cooperation,Broom1997Multiplayer,Pena2014Gains,Newman2003Structure,
Li2015Evolutionary,Zhang2013Tale,Gokhale2014Evolutionary}. However, with the increase in internet-based interactions (such as social media, self-organized collaborative communities, sharing economy etc.) graph theory seems to re-emerge under the headline of social networks, now with the additional advantage of a new empirical contribution of large amounts of data. Comparing the system behaviour in the fully connected graph (i.e. mean field model or \textit{panmixia}) with more realistic spatial interaction models, including those with population structure, can reveal the effect of the aforementioned simplifying assumptions on the level of cooperation, see e.g. \cite{Ohtsuki2006Simple,Szolnoki2012Conditional}.

 In this paper we study perhaps the simplest version of a true multi-player dynamic: the threshold game (this claim is elaborated on below). Inspired by \cite{Bach2006Evolution}, we focus on $N$-player games, in which $M$ out of $N$ participants have to decide to cooperate for anyone to receive a reward. Thus, $N$ is the number of people interacting in a given match, and $M$ is the threshold, describing the minimal number of cooperators necessary for pay-off to take place. Additionally, we follow \cite{Santos2012Dynamics} by placing the population on a network, in which the $N$-sized groups are picked based on the connections between individuals. This is inspired by the fact that in everyday life in both animal and human systems, an individual may participate in several different groups, with either a high or low degree of overlapping members. In either case, it is quite possible that the experiences gained in one setting are applied when deciding upon a strategy in another.
  Note that we have not implemented any kind of load-sharing, meaning that cooperation has the same cost irrespective of the number of cooperators. It is worth noting that we still recover the qualitative results shown in, e.g, \cite{Zheng2007Cooperative}, where cost sharing is included.

 For those who are thus inclined, one can visualize each $N$-player match as a party in which a certain number of guests have to volunteer for food preparation for it to be done in time. If too few people volunteer, the food is never ready and no one gets anything, hence the threshold. If too many people volunteer, chaos ensues, reducing efficiency, hence the lack of load sharing (i.e. the cost of cooperation does not go down with increasing cooperators). The precise and more rigorous description of the model is given in the next section.

\section{Model}
 
\begin{figure}
\centering
\includegraphics[width=0.45\linewidth]{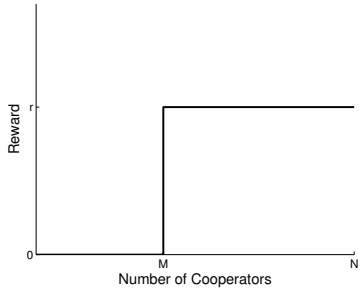}
\caption{The reward for participating in a game with $\#C$ cooperators.}
\label{fig:nonlinearSchematic}
\end{figure}

The game has been chosen as the simplest truly multiplayer game. The criterion for being a true multiplayer game is that it should not be possible to describe the payoff from a match as a sum of two-player interactions; i.e. the payoff from a match should behave nonlinearly as a function of cooperators \cite{Li2015Evolutionary,Zhang2013Tale}. We have chosen to focus on this property because we expect that a linear payoff-dependence in an N-player game would be equivalent to a 2-player game in which fitness was calculated as an average over all interactions. As the explicit target of this investigation is N-player dynamics, we believe nonlinearity to be crucial. Possibly the simplest, interesting functional form meeting this nonlinearity requirement is a step function (Fig \ref{fig:nonlinearSchematic}). It is worthwhile to note here that \cite{Archetti2011Coexistence} has shown convincingly that all multiplayer games with payoff structures depending nonlinearly on the number of cooperators have attractors qualitatively similar to what is studied here; i.e. the step function is a suitable starting point for a more general investigation. This also indicates that that this model has the potential to exhibit behaviour different from what has already been extensively described for pair-wise models \cite{Wang2015Universal}. We think that this finding by \cite{Archetti2011Coexistence} supports our intuition regarding nonlinearity very well.

 Thus, the game studied here is a threshold game in which payoff only takes place when the number of cooperators reaches a certain threshold, $M$. One could envision the dinner party described above, or group hunting carnivores where the prey is only brought down when a certain number of individuals participate, and where the cost, in the form of energy spent on running, depends on prey rather than participants. On another level one could imagine a collection of interacting cells or bacteria producing growth factors or drug resistance compounds, and only if a sufficient number of cellular units contribute would the common good be manifested \cite{Bach2001Evolutionarygame,Hummert2014Evolutionary}. 

We consider a finite population of $L$ individuals, each with the choice of either cooperating (C) or defecting (D). The $i$'th individual is represented by her probability, $x_i$, of cooperating. $x_i$ is called the "strategy" of the $i$'th individual. The individuals participate in $N$-player games, in which each participant of a given match is rewarded $r$ if at least $M$ players cooperated, and none otherwise. In this way it is a simplification of the game described in \cite{Zhang2015How}. Irrespective of the number of cooperators, the cost of cooperating for a given individual is $c$. The game is a generalization of other games in which multiple players must coordinate their behaviour. For $N=2, M=1$, this game is reminiscent of the classic snowdrift or hawk-dove game \cite{Smith1973Logic}. For $M=1$, it is the volunteers dilemma \cite{Weesie1993Asymmetry}.

The simulation proceeds in a number of rounds. At the beginning of each round, all individuals decide on playing either C or D, according to their respective strategies. During the course of a round, $G$ matches are played, in each of which $N$ individuals are picked from the population, and their shared payoff according to the rules of the game is calculated before returning them to the pool. The manner in which the players are matched is described below. At the end of a round, each individual calculates the average payoff per match of all C-players and all D-players that the individual has encountered during the round, $f_C,f_D$. We assume that the player has full knowledge of the fortunes of these players, irrespective of how many of the $G$ matches they actually shared. The strategies of the individuals are then updated as follows:

\begin{equation}
x_i \rightarrow x_i(1-x_i)(f_C-f_D)dt , \label{eq:update}
\end{equation}

\noindent where $dt$ is a discretization time step, equal to $0.01$, while $f_C,f_D$ are generally of order 1. The update takes place after each round of $G$ matches, such that these are in effect simultaneous. If a player does not meet both cooperators and defectors during a round, then $f_C$ (or $f_D$) has the same value as in the previous round.

 All simulations are run for $10^5$ iterations, such that the amount of simulated time is $10^5dt=1000$. At the beginning of the simulation, the $x_i$ are drawn from a uniform distribution on $\left[0,1\right]$.
 
 Simulations were run in Matlab, and extensive summaries of the results have been made available as a mySQL database on Dryad.

\subsection{Mean Field Treatment}
\label{sec:meanfield}
\noindent The model described thus far has been studied with a mean field approach in \cite{Bach2006Evolution}. There it was found that there are up to 4 solutions of $\frac{d\meanx}{dt}=0$ in the mean field model: $\meanx=1,0$ and the roots of 

\begin{equation}
g(x)=r\binom{N-1}{M-1}\meanx^{M-1}(1-\meanx)^{N-M}-c \qquad .
\label{eq:meanfield}
\end{equation}
Note that \meanx in this paper will refer to the population average of strategies, evaluated instantaneously.
The lower root of (\ref{eq:meanfield}) and 1 are repellers – any deviation however small will lead the system away from the solution, while 0 and the upper root are attractors, or evolutionary stable states, as they are also called. These circumstances turn out to be important to our findings, which will be discussed below.

Defining the parameter $\alpha=c/r$, i.e. the ratio between cost and reward of cooperation, it was furthermore shown in \cite{Bach2006Evolution} that (\ref{eq:meanfield}) only has real roots when 
\begin{equation}
\alpha \leq \binom{N-1}{M-1} \left( \frac{M-1}{N-1} \right)^{M-1}\left(\frac{N-M}{N-1}\right)^{N-M} \quad .
\end{equation}
Because of this, we will henceforth use $\alpha$ as the principal means of describing $r,c$-variation.

\subsection{Population structure}
\begin{figure}%
\centering
\subfloat[][ Fully mixed network with no stable connections.]{\includegraphics[width=.45\linewidth]{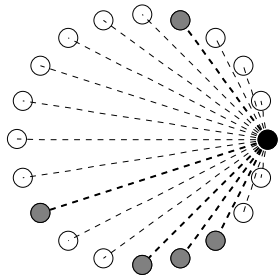}
 \label{fig:noTopo}}%
\qquad
\subfloat[][Random network.]{\includegraphics[width=.45\linewidth]{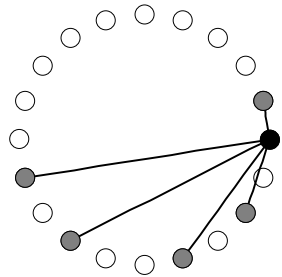}
\label{fig:randomGraph}}
\qquad
\subfloat[][Ring-like network with no long-range connections. ]{\includegraphics[width=.45\linewidth]{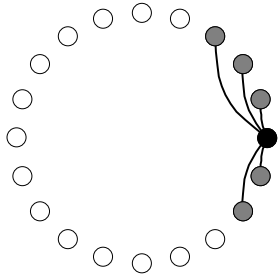}
 \label{fig:noLongGraph}}%
\qquad
\subfloat[][Ring-like network with long-range connections. For odd $N$, two long connections are made, to preserve symmetry.]{\includegraphics[width=.45\linewidth]{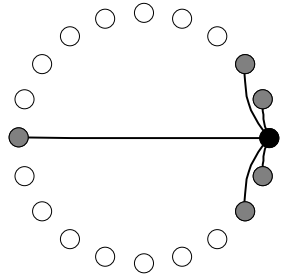}
 \label{fig:1Long}}%
\caption{Depictions of the different network types studied. In each case the connections of a single individual are highlighted. $L=20, N=6$.}%
\label{fig:cont}%
\end{figure}

\begin{figure}
\centering
\includegraphics[width=0.45\linewidth]{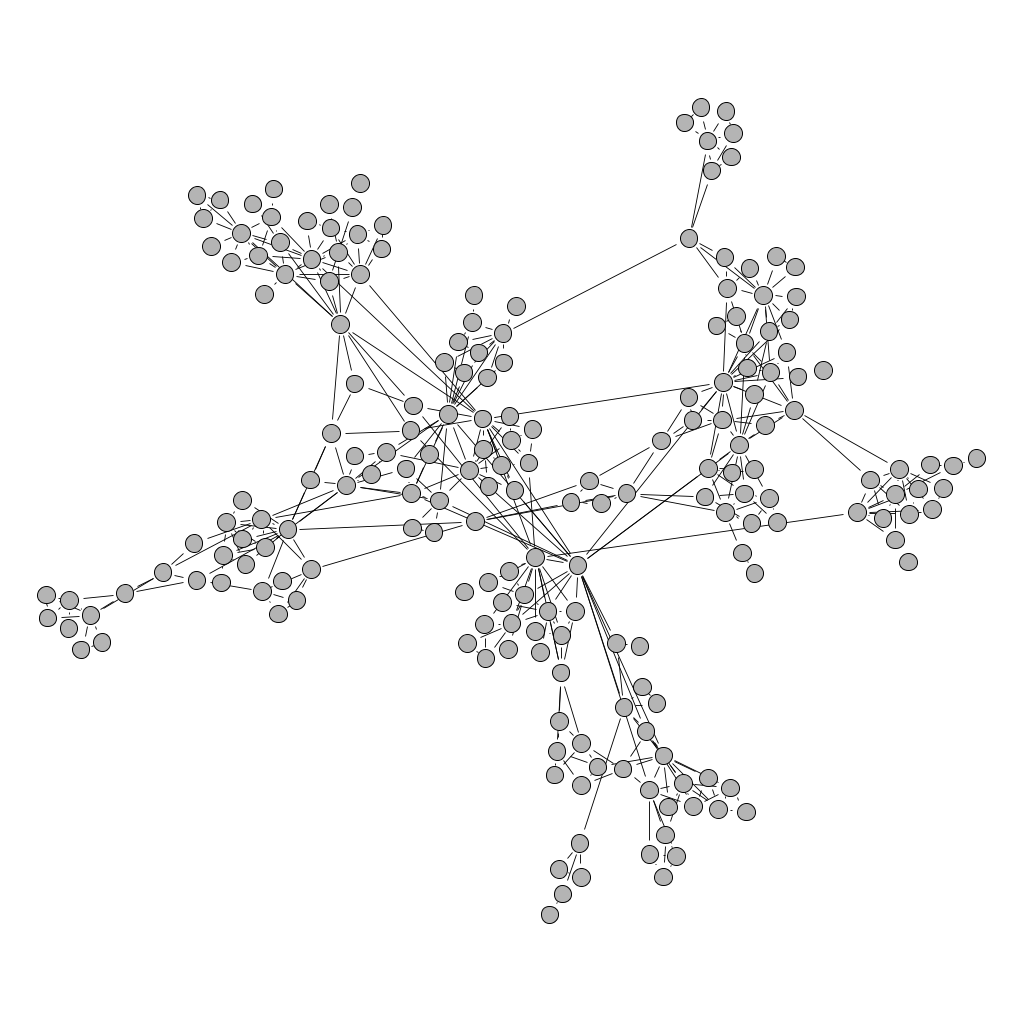}
\caption{An example of a "social" network, as described in \cite{Toivonen2006Model}, for $L=300,f_1=f_2=0$. See the appendix for further details on the generation and $f_1,f_2$. All connections between individuals are drawn. Proximity in space reflects connectedness. Plot created using Gephi \cite{ICWSM09154}.}
\label{fig:socialNet}
\end{figure}

\noindent We implement population structure by requiring that an individual is only matched with people with whom they are connected (befriended). For all networks, $G=L$, and, in all but the first network type, each group of $N$ is picked as the $i$'th individual and its $N-1$ neighbours, for $1\leq i \leq L$. This matching rule is similar to what was used in \cite{Santos2012Dynamics}.  We consider five different types of population structures, or networks:

\begin{description}
  \item[Fully mixed:] This is the standard case, in which the $N$ players for each match are picked randomly from the entire population. In network terms, we may think of this as the "fully connected" case. As such, it is the only network type which diverges in its matching rule, in that an individual will not necessarily be matched with all her connections in a given round. The situation is depicted in Fig. \ref{fig:noTopo}.
  \item[Random, regular network:] In the random, regular network, the population is placed on an undirected, regular graph, each vertex representing an individual. The degree of each vertex is $N-1$. An example is given in Fig. \ref{fig:randomGraph}.
  \item[Ringlike, regular network, short range connections:] In this network, we imagine the individuals placed on a circle. If $N-1$ is even, each individual is connected to the $N-1$ individuals closest on the circle. If $N-1$ is odd, the individual is connected to the $N-2$ closest neighbours, and the last connection is made to the first non-connected individual in either clock- or counter clockwise direction. It is attempted to alternate between the two directions. It is, unfortunately, not possible to create this network for all combinations of $N$ and $L$ if $N$ is even, which may be noticed by careful examinations of the following figures. An example of the layout is given in Fig \ref{fig:noLongGraph}.
  \item[Ringlike, regular, added long range connections:] Very similar to the previous type, but with added connections to the opposite side of the ring. If N is even, one long-range connection is made, if odd, two are, to avoid the above-mentioned problems concerning even $N$. It is worth noting that these networks are closely related to the small-world networks studied in \cite{Watts1998Collective}. An example is seen in Fig \ref{fig:1Long}.
  \item[Social Network:] As an attempt at a realistic case, we follow the algorithm of \cite{Toivonen2006Model} to synthesize networks with high clustering and varying degree. As described in the appendix, the two attachment probabilities can be adjusted to obtain networks with a wide range of structures. An example plot is shown in Fig \ref{fig:socialNet}.
\end{description}

\noindent Unless otherwise stated, the size of the populations are $L=300$. We have made this choice after verifying that \meanx as a function of $\alpha$ for $L=600$ is indistinguishable from the case $L=200$, meaning that any $L$ above 200 is most likely sufficient for the ranges of $N,M$ studied here. All reported values of \meanx are the result of averaging over at least 5 different initial conditions.

\section{Results and discussion}

\subsection{Requirements on $N$}

Somewhat surprisingly, we find that population behaviour does not become representative of general multiplayer games until $N$ is at least 4, and not 3 as would have been naively expected. An example of this behaviour is given as part of the discussion of \meanx vs $\alpha$, and is, we suspect, due to the fact that for $N=3$ the only permitted network structure is a collection of rings, which are very different from all other networks here studied. Because of this, we have decided to focus on $N>3$ in the remainder of the paper.

\subsection{Regular networks}

As was shown in \cite{Bach2006Evolution}, in the mean field model, there is an abrupt transition between a cooperative regime in which two collective states are possible, a mixed and a fully defecting, and a regime in which only the fully defecting state is possible. This transition takes place as the cost-to-reward fraction, $\alpha$, changes. As will be detailed below, we reproduce this qualitative behaviour for regular networks, with some adjustments depending on the details of the network. To simplify things, we will often focus exclusively on the critical $\alpha$ value, $\alpha_{crit}$, at which the transition takes place. To keep things simple, we define $\alpha_{crit}$ in a numerical model as the $\alpha$-value for which $\meanx \cdot N/M$ crosses 0.1, and in the mean field model as the point where the mixed state disappears (i.e. where the discriminant of (\ref{eq:meanfield}) becomes negative). The choice of $0.1$ as the point of transition stems from the observed shapes of \meanx vs. $\alpha$, see later figures for examples.

\subsubsection{$N$-dependence}

\begin{figure}[htbp]%
\centering
\includegraphics[scale=.55]{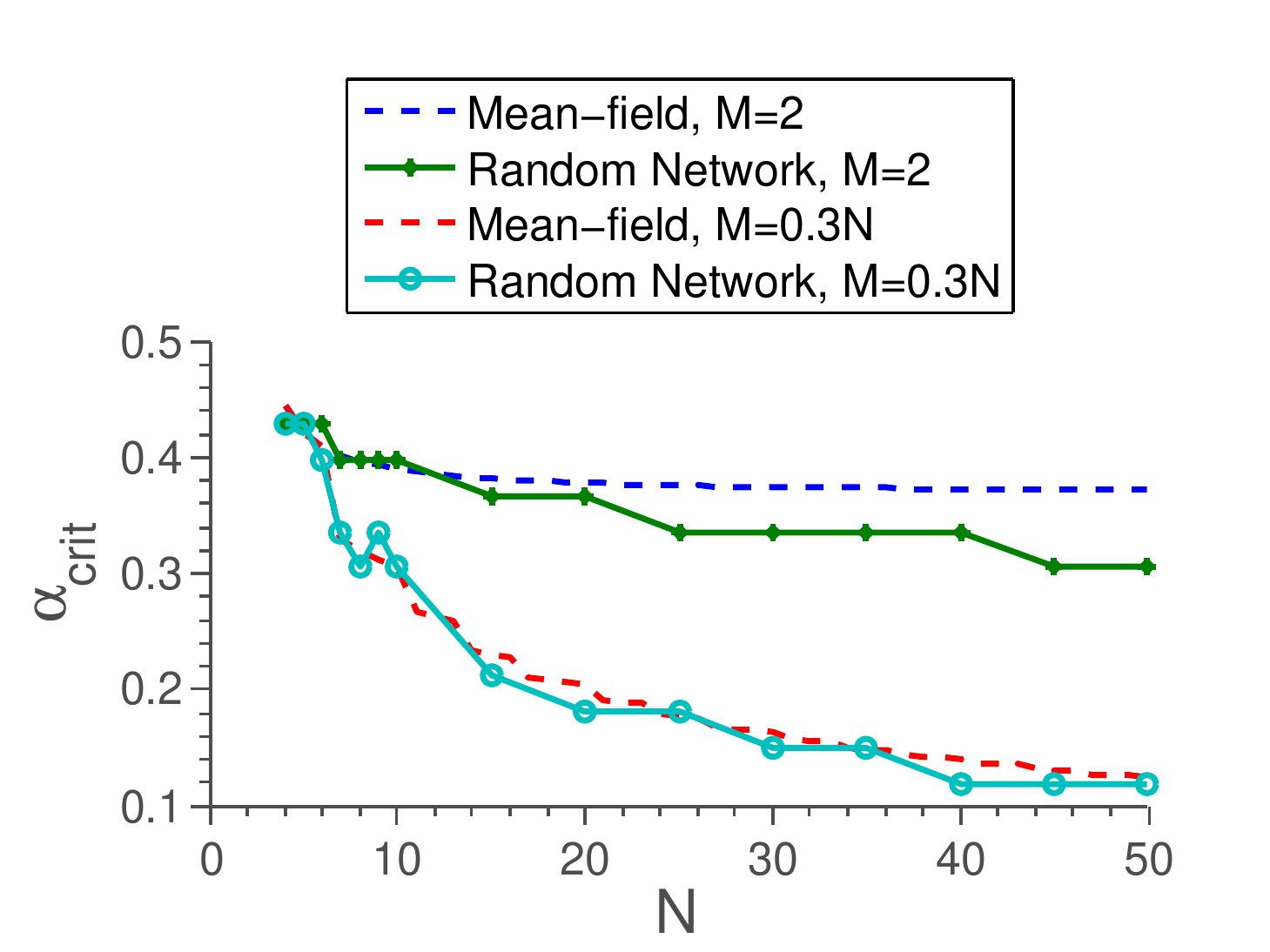}
\caption{ $\alpha_{crit}$, both predicted from the mean field theory as well as measured from the model implemented on regular, random networks. We see that for both static $M$ as well as relative, there is an $N$-dependence. However, in the static case, when $M<<N$ the dependence is not predicted by the mean field theory. }%
\label{fig:Ndepend}%
\end{figure}

\begin{figure}%
\centering
\includegraphics[width=.45\linewidth]{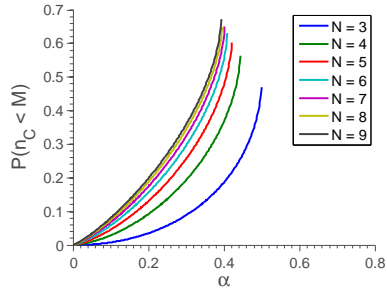} 
\caption{The probability of having insufficient cooperators as a function of $N$, for $M=2$, assuming that \meanx is equal to the upper root of (\ref{eq:meanfield}), as predicted by the mean field theory. Note that the different extents of the lines (i.e. that $N=9$ covers a smaller part of the $\alpha$-axis than $N=3$) stem from the limitations of the mean field model.}
\label{fig:mfPred}%
\end{figure}

We have studied $N$-dependence in two cases; $M$ static, and $M/N$ constant. In the static case, we focus on $M=2$ unless otherwise stated. In Fig \ref{fig:Ndepend} is shown $\alpha_{crit}$ for both setups, and we see an $N$-dependence in both cases. In the case of the static $M$, this is somewhat surprising, since it means that even though more individuals are available to solve the same problem, the chances of a sufficient amount of cooperators decrease. We can understand this low-$M$ behaviour by taking into consideration that \meanx will tend to decrease as a function of $N$, as also predicted by the mean field model (it is beneficial for the individual to do less when more people are there to share the burden), combined with the fact that the number of cooperators in any given match is stochastic:

Let $n_C$ be the number of cooperators in a given game. For a proportion $\left\langle x\right\rangle$ of C-players of the entire population, evenly distributed across the graph, the distribution of $n_C$ is given by a Bernstein polynomial of degree $N$:

\begin{gather}
P(n_C=k)=\binom{N}{k} \left\langle x\right\rangle^k (1-\meanx)^{N-k}  
\Rightarrow
P(n_C<M)=\sum_{k=0}^{M-1}\binom{N}{k} \meanx^k (1-\meanx)^{N-k} \ .
\end{gather}
We can now substitute \meanx with the predicted value of \meanx for the mean field model (upper root of (\ref{eq:meanfield})). Fig \ref{fig:mfPred} shows the result of this, and we see that the exercise predicts the probability of $n_C<M$ in a given game to increase for increasing $N$. Taking into consideration that the lower root of (\ref{eq:meanfield}) is repelling, and that full defection ($\meanx=0$) is a completely stable state (there will be no fluctuations in $n_C$), we find that the $N$-dependence seen in Fig \ref{fig:Ndepend} can be understood as arising from the fluctuations in $n_C$ in the mixed state. This conclusion is further supported by the fact that when this  fluctuation analysis is repeated for constant $M/N$, a similar breakdown is not predicted for large $N$. Indeed, it is predicted that $P(n_C < M)$ decreases. This matches our observations from the simulations.

This effect resembles what was reported in \cite{Archetti2009Volunteers}, in that it reports decreasing likelihood of the population as a whole to meet the threshold, for increasing $N$. In more general terms, it is reminiscent of the bystander effect \cite{Latane1972Unresponsive}: the chance of the necessary number of cooperators appearing \textit{decreases} for an increasing number of participants. A similar effect was reported in \cite{Chen2010Effects}.

\subsubsection{Topology Dependence}

\begin{figure}%
\centering
\subfloat[][ Example behaviour of mean field, full mixing and random networks, $N=7$.]{\includegraphics[width=.45\linewidth]{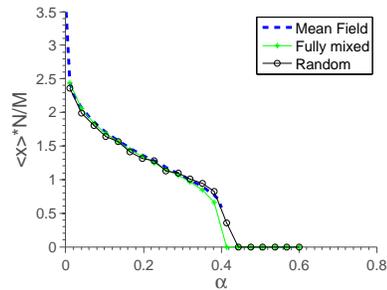} 
\label{fig:meanFieldComparison2}}%
\qquad
\subfloat[][Example behaviour of rings without long connections. The outlier is $N=3$.]{\includegraphics[width=.45\linewidth]{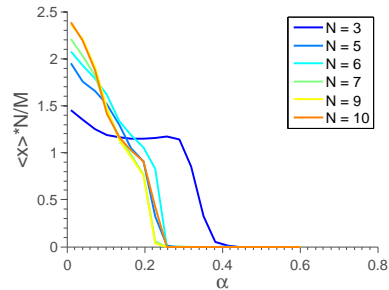}
\label{fig:no_long_N3}}
\qquad
\subfloat[][Example behaviour of rings with few long connections]{\includegraphics[width=.45\linewidth]{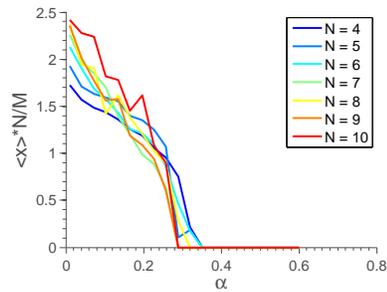}
}%
\qquad
\subfloat[][Comparisons of critical $\alpha$-values for different models, as a function of $N$. "Fully Mixed" and "Random Topology" overlap for part of the range.]{\includegraphics[width=.45\linewidth]{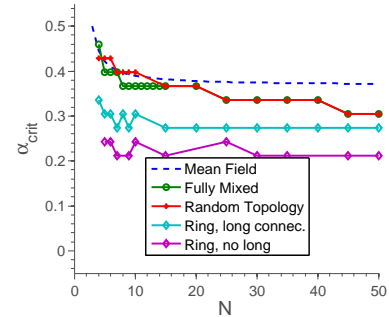}
}%
\caption{(Color online) Population average of $x$ for various systems with threshold dynamics, for $M=2,dt=0.01, L=300$. Note that in a-c, the value plotted is $\frac{N\meanx}{M}$, such that a value of 1 means that on average, exactly $M$ players are cooperating in each match.}%
\label{fig:observations}%
\end{figure}

In Fig \ref{fig:observations} is shown the behaviour of \meanx for different network topologies, as described in Fig \ref{fig:cont}. We find that while the qualitative behaviour of \meanx is the same, the details depend on  the topology of the network. We find that $\alpha_{crit}$ decreases when the network is wired such that the connections become as local as possible. This result is closely related to that reported by \cite{Hauert2004Spatial}, who found that adding spatial structure to a population reduces cooperation in two-person snow drift games. However, it is worth noting that what is demonstrated here is a considerably more general result, in that we here go beyond dyadic interactions, and grid-based populations.

\begin{figure}[htbp]
\centering
\subfloat[][ Comparison of finite size effects to clustering effects]{\includegraphics[width=.45\linewidth]{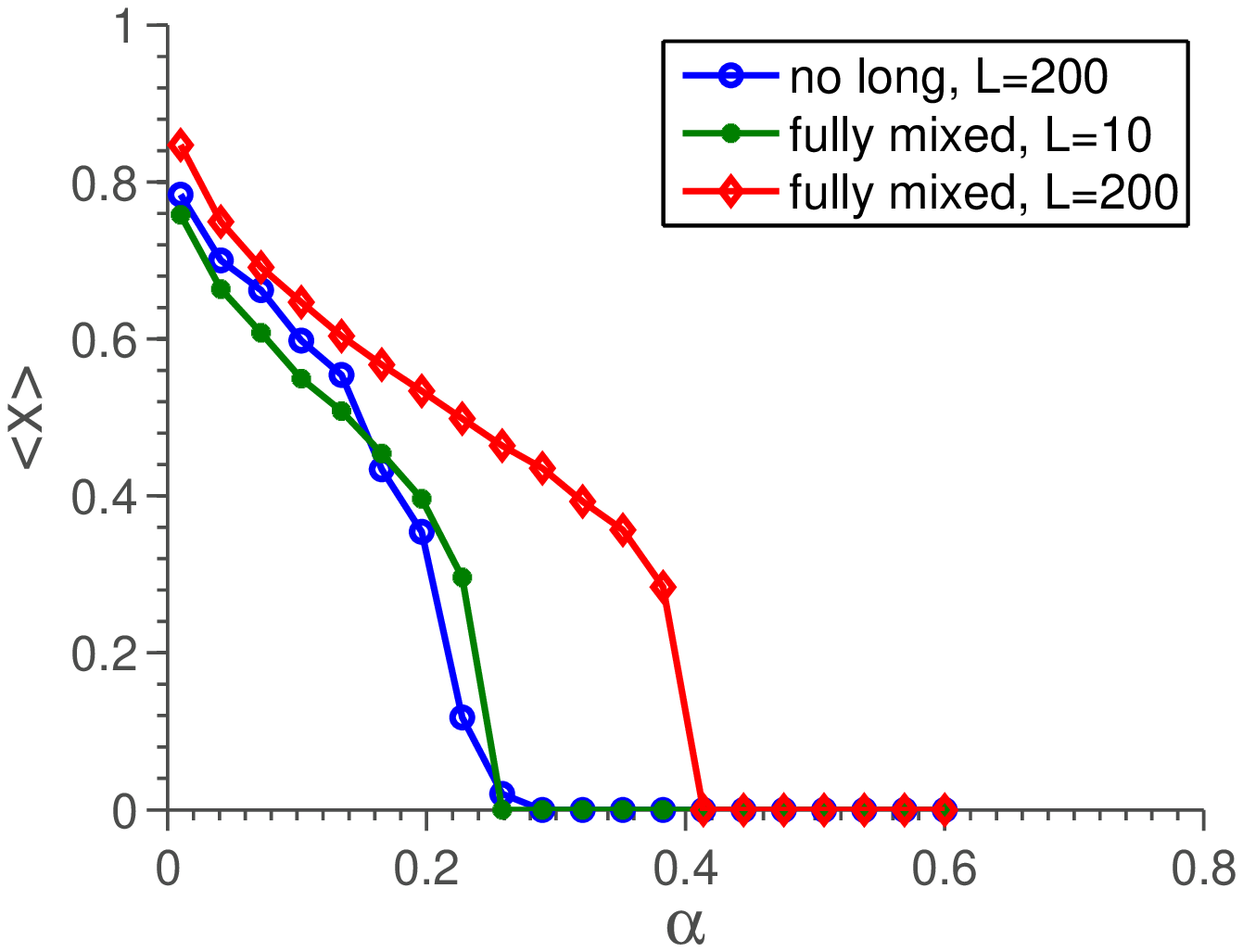}
 \label{fig:effL_visual}}%
\qquad
\subfloat[][Scatter-plot of $L_{eff}$ vs. $L^*$ (defined in (\ref{eq:Ltilde})). We see an approximatively linear relation.]{\includegraphics[width=.45\linewidth]{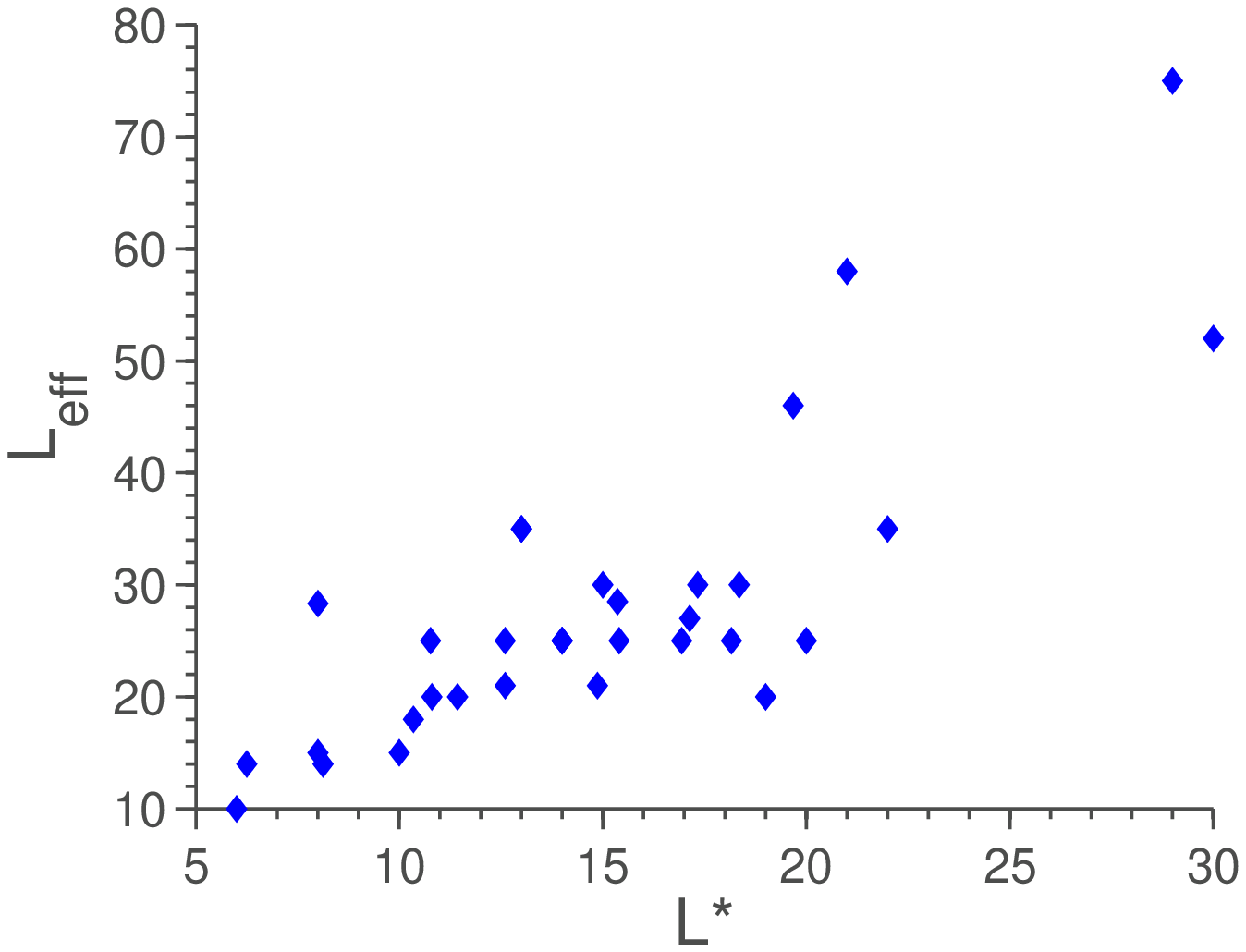} 
\label{fig:effL_comp}}
\caption{Evaluation of the finite-size explanation of the topology dependence in \meanx vs. $\alpha$. }%
\label{fig:effL}%
\end{figure}

We propose that this topology dependence comes about through the way in which each individual experiences the rest of the population - that in fact clustering leads to the individuals behaving as if the population had an effective population size\footnote{Please note that, despite the name, $L_{eff}$ is not directly relatable to the similar term in population genetics, as also pointed out in the conclusion, below.}, $L_{eff} < L$, making the topology-effect a finite-size effect induced by clustering. The first observation to make in this regard is Fig. \ref{fig:effL_visual}, where we see \meanx vs. $\alpha$ for both a network with only short-range connections but large $L$, and a fully mixed small-$L$ network. We see here both that indeed small populations are prone to smaller $\alpha_{crit}$, but also that for a given $\alpha_{crit}$ and $N$, an $L$ can be found such that the fully mixed model matches. This matching works as our definition of $L_{eff}$. It is in this connection worth noticing that by defining the $L_{eff}$ based on comparisons to networks with the same $M$ and $N$, any variation in $L_{eff}$ must be due to effects other than the $N$-dependence already discussed.

 Considering the cause of $L_{eff}$, we may consider the number of individuals that a given player can meaningfully be said to interact with. We know that the individual interacts with the immediate neighbours, of which there are $N-1$. However, due to the matching rules as laid out in the previous section, the player also interacts with each of their neighbours (discounting herself), of which there are almost $(N-1)^2$. Of these, $\gamma\cdot (N-1)^2$ are duplicates, $\gamma$ being the clustering coefficient of the network. In short, we may imagine the number of direct and indirect interactions of an individual to be on the form
 
 \begin{equation}
 \tilde{L}\approx A_0(N-1)+A_1(1-\gamma)(N-1)^2+A_2(1-\gamma)^2(N-1)^3 + ... \quad ,
\label{eq:L}
 \end{equation}
where the $A_i$ are weights used to signify that the interactions with neighbour's neighbours and so on are weaker than with direct neighbours. As (\ref{eq:L}) is too complicated for a direct comparison to $L_{eff}$, we have instead tried to use the same basic intuition behind (\ref{eq:L}) to propose a much simpler expression:

\begin{equation}
L^*=\frac{I}{1+\gamma} \quad , I: \text{ number of neighbours \& neighbours' neighbours}
\label{eq:Ltilde}
\end{equation}

Here $I$ is a function of both $\left\langle N \right\rangle$ and $\left\langle \gamma\right\rangle$, and the denominator is there to provide additional dependence on $\gamma$, since the effect appears to be marked. Using (\ref{eq:Ltilde}) we may test our basic intuition about the system by checking the relationship between $L_{eff}$ and $L^*$. Looking to Fig. \ref{fig:effL_comp}, we see that indeed, $L_{eff}$ is well correlated with $L^*$ (correlation coeff.: 0.7). We interpret this to mean that our qualitative explanation of how topology influences cooperation is correct. As to why small $L$ leads to low cooperation, it seems plausible that it is related to small populations being more susceptible to fluctuations in \meanx, which will be much more pronounced in the mixed state than in the $\meanx=0$ state. In other words, that the observed decrease in cooperation is driven by the differences in $\meanx$-fluctuations within the two evolutionary stable states, like the case was for the above discussed $N$-dependence.

\subsection{Social Networks}
\noindent An interesting question is to what extent the findings on regular networks generalize to the arguably more relevant case of non-regular networks, such as that depicted in Fig \ref{fig:socialNet}. To accommodate the spread in the number of players in a given match on such a network, we have here decided to use a relative $M$, such that $M=\ceil{N\cdot M_{rel}}$, with $M_{rel}$ being a number between 0 and 1. This choice was made because most choices of $M>1$ would result in many matches having $N<M$, for networks such as the one shown in Fig \ref{fig:socialNet}.

\begin{figure}%
\centering
\subfloat[][ ]{\includegraphics[width=.45\linewidth]{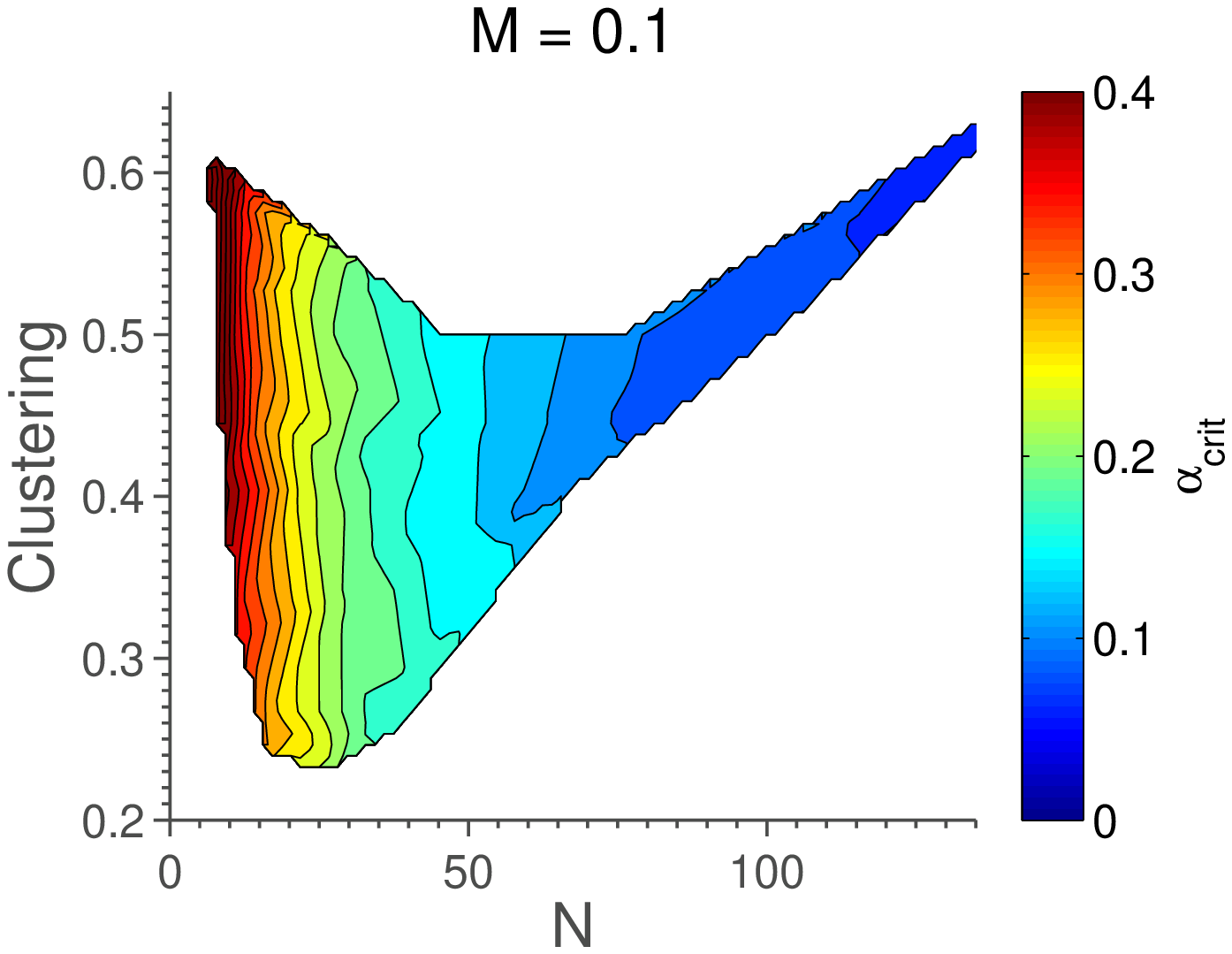}
}%
\qquad
\subfloat[][]{\includegraphics[width=.45\linewidth]{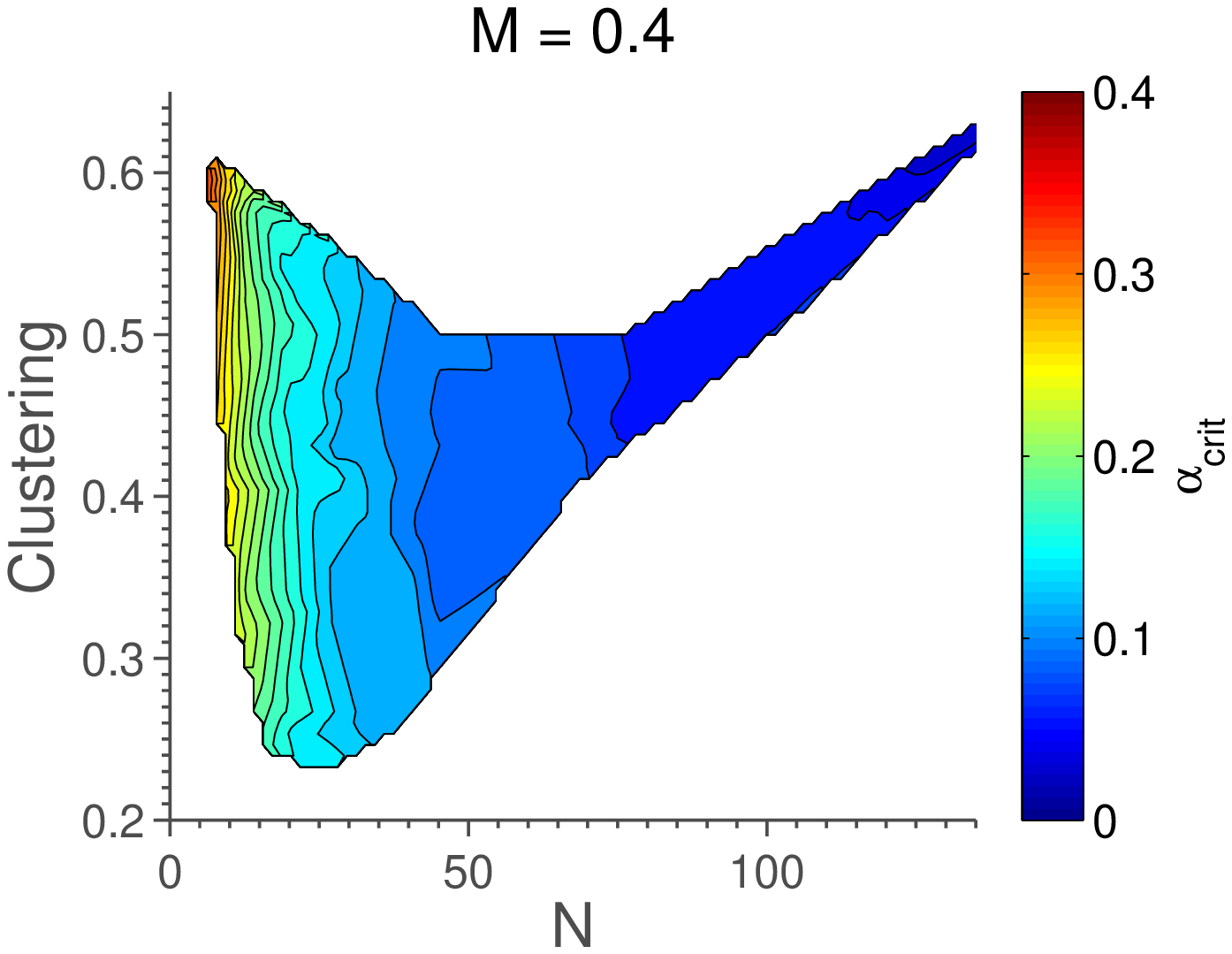}
}
\qquad
\subfloat[][]{\includegraphics[width=.45\linewidth]{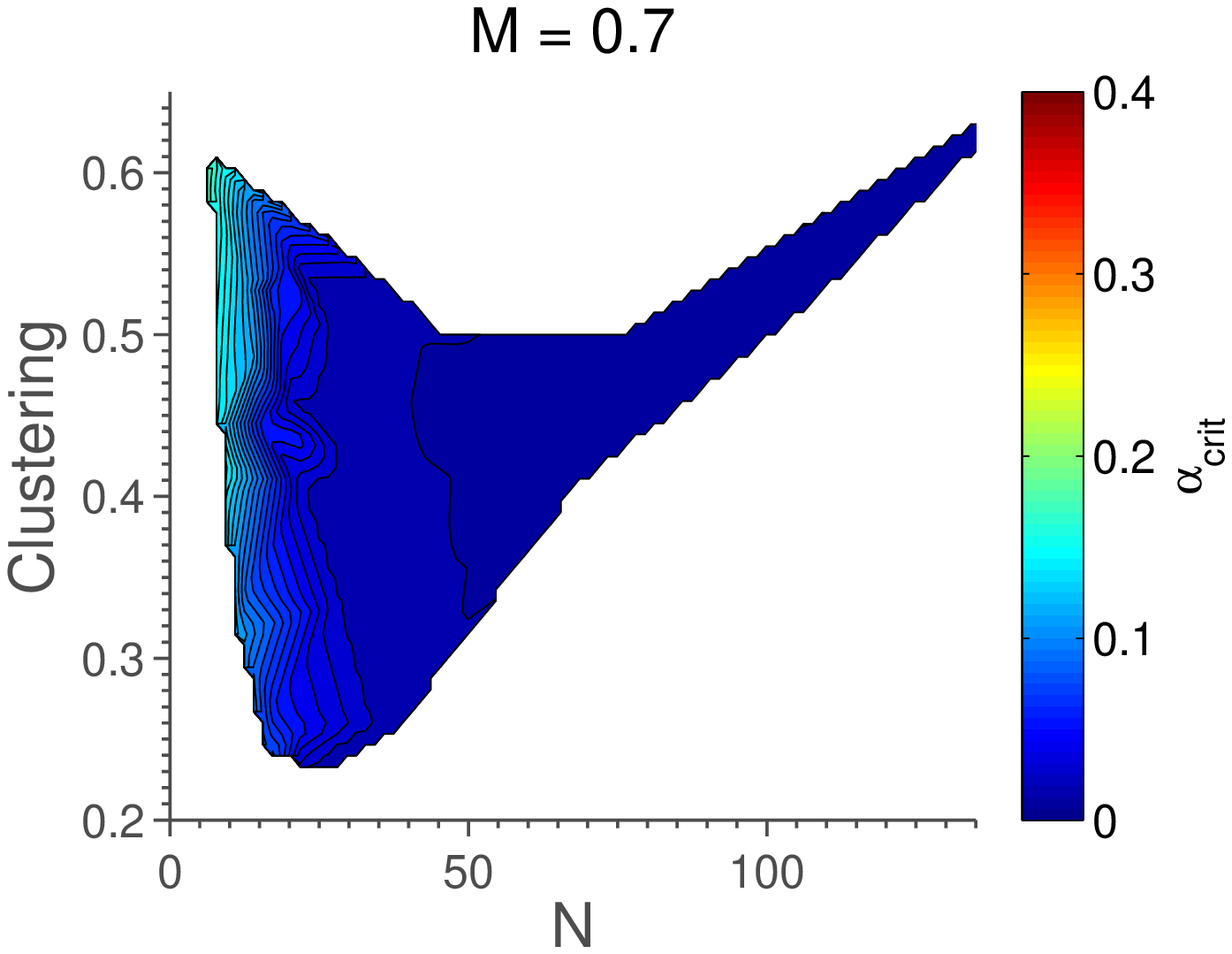}
}%
\caption{Measured $\alpha_{crit}$ for various realizations of the social networks. We see that in particular the $N$-based variation is strong. The coloured region coresponds to the $(0,15) \times (0,35)$-region in $(f_1,f_2)$-space; see the appendix for an explanation of $f_1,f_2$.}%
\label{fig:socialNet_crit}%
\end{figure}

 In Fig \ref{fig:socialNet_crit} is shown $\alpha_{crit}$ as a function of $\left\langle \gamma\right\rangle$ and $\left\langle N \right\rangle$ (defined as average number of neighbours + 1). We see that a range of $\alpha_{crit}$ is possible, and also that the primary cause of variation seems to be the $N$-dependence discussed above. Note that the peculiar shape of the coloured region is due to the fact that we are unable to sample the $(\left\langle N \right\rangle,\gamma)$-space directly, but are instead exploring the $(f_1,f_2)$-space, for which $(0,15) \times (0,35)$ maps to the depicted region. See the appendix for an explanation of $f_1,f_2$.

\subsection{General Behavior}
\noindent We see that all studied deviations from the mean field model appears to reduce cooperation, in particular $\alpha_{crit}$, to some extent. This can be understood by considering the attractors in the mean field model:

As was discussed in Sec. \ref{sec:meanfield}, the mean field model has been shown to have, for $\alpha < \alpha_{crit}$, two attractors and two repellers in \meanx-space. What this means is that if the system is brought to cross the repeller dividing the two attractors (the lower root of (\ref{eq:meanfield}), the dynamics of the mean field model will dictate a transition from one state to the other. What the mean field model does not include is that the actual number of cooperators in a given match is stochastic, except when \meanx is 0 or 1. This leads to a certain noisiness in the simulation results, or fluctuations in \meanx as perceived by the individual, in the mixed state, but not in the $\meanx=0$ state.

Hence, since all perturbations to the initial model, studied here, have had the consequence of exacerbating the effects of the fluctuations, a higher rate of transition from mixed state to fully-defective state is observed, most critically close to $\alpha_{crit}$, where the downwards jump in \meanx necessary to clear the repeller is smallest. A consequence of this mechanism is that we would expect similar dependence on both $M/N$ and clustering for any game where two attractive states had this difference in inherent fluctuations. We note that, as shown in \cite{Bach2006Evolution}, the existence of the mixed state requires $N>2$ for the threshold game. This requirement goes a long way to explain why the structure-related behaviour described here has not already been extensively described in the literature, which has primarily been focused on 2-player games.

\section{Conclusion}

In this paper we study the simplest truly multiplayer game, the threshold game, in structured populations. We find that the average behaviour of the players as a function of the cost-to-reward fraction, $\alpha$, is highly dependent on the topology of the network describing the population, both in terms of number of neighbours as well as higher level effects such as clustering. We also find that structure appears to primarily decrease cooperation, by destabilizing the mixed pseudo-equilibrium in favour of the fully defecting state; at least to the extent that the resulting system can be compared to the mean field or fully mixed models. The observation that spatial structure
can inhibit the evolution of cooperation was made earlier for the more restrictive case of dyadic interactions and only for a two-dimensional grid implementation of space exclusively with nearest neighbour interactions \cite{Hauert2004Spatial}. Here we substantiate the speculation that this evolutionary behaviour is observed in a more general context, i.e. $N$-player scenarios and network structures representing realistic social networks.

We find that, irrespective of the threshold for payoff, larger numbers of players in each match ($N$) result in less cooperation. In particular we find that for $M/N$ very small, meaning a relatively low threshold ($M$), the stochastic nature of the game destabilizes cooperation further. Given the good correspondence between the theoretical predictions and numerical observations in this paper, we suggest that related scenarios (with payoff-functions that can be approximated by a step-function) give rise to similar evolutionary dynamics.   

 Additionally, we find that the behaviour on highly clustered networks is similar to the behaviour observed in very small, fully mixed, populations, leading us to suggest a working notion of an effective population size, $L_{eff}$, to describe the behaviour for a given population structure with respect to the transition. It is shown in Fig \ref{fig:effL_comp} that $L_{eff}$ to a large extent can be predicted based solely on the degree and average clustering of each vertex in the network. It should be mentioned that this effective network population size bears little resemblance to the family of well known genetic effective population sizes \cite{Wright1938Size,Ewens1982Concept,Gillespie2001Population}, except maybe the panmixia assumption. 
 
 It is interesting to note that we do not seem to echo the findings of \cite{Santos2006Evolutionary}, who, by studying pair-wise interactions on graphs, found a marked increase in cooperation for heterogeneous networks. While a direct comparison is not possible in this study (since the rules of the game, and hence the expected results, depend on the degree of the vertex), Fig. \ref{fig:socialNet_crit} does not lead us to believe that a similar effect is at play here. Presumably this, like most of our other findings, is due to the bistable nature of the threshold game.

The results presented in this paper show, perhaps not surprisingly, that within multiplayer games on structured populations, there is an intricate interplay between the details of the game and the structure of the population. However, it also bears noting that much of the behaviour observed can be seen to stem from properties of the mean field model - especially regarding the relative stability of the mixed and full-defection states. It is important to point out that as was shown in \cite{Bach2006Evolution} and \cite{Archetti2011Coexistence}, the nonlinearity of the payoff function as well as the larger number of players ($N>2$) are both necessary requirements for the existence of the mixed state. As such, models lacking these features would not be expected to have a similar dependence on population structure.

Finally, it is worthwhile to remember that while the present model is surely quite simple - it has no reputation, no kin-selection, no outside forcing - all refined models taking these concepts into account, but retaining population structure and payoff non-linearity, will most likely exhibit behaviour related to, or at least moulded by, the mechanisms uncovered in this paper. As such, this simplified model is relevant for all such other more specialized takes on the subject of cooperation in structured populations, as was also argued in \cite{Archetti2011Coexistence}. 

An example of a slightly more advanced model is given in the appendix, where players are allowed to also make suboptimal choices in their update strategies. It is found that the general results are still valid. Furthermore, it is relevant to note that \cite{Santos2012Dynamics} uses slightly different payoff structures and update rules, but still present results showing that the cooperative (mixed) state becomes less stable when structure is introduced.

\section{Acknowledgements}
This project has been supported by the seed funding program from the Interacting Minds Centre, Aarhus University. Furthermore, the authors are grateful for valuable feedback from Andreas Roepstorff, Chris and Uta Frith as well as our anonymous reviewers.

\appendix

\section{Network Generation}
\subsection{Regular networks}
When generating regular networks, we have chosen the quite straightforward algorithm of always starting with a ring-like network with an even number of close connections. The connection matrix for this case is trivial to create, based on the diagram in Fig \ref{fig:noLongGraph}. If an additional short-range connection had to be added to this, we would go about it in the following systematic manner:
\begin{enumerate}
\item pick 1 vertex on the circle
\item If it is not of sufficient degree, connect it to the first non-neighbour in clockwise / counter clockwise direction.
\item Proceed to the next vertex in the clockwise direction.
\end{enumerate}

We would then repeat steps 2 $\&$ 3 alternating the clockwise / counter clockwise decision in step 2. As has already been mentioned, this procedure is not guaranteed to work for all $N$, even if $L$ is required to be even. We have chosen not to study those sets of $L,N$ where the above method did not succeed.

When a random, regular network was needed, we have created a ring-like network, if needed with a single long-range connection, depending on the desired degree. To then obtain a random network, it is sufficient to repeatedly pick 2 edges at random: a--b $\&$ c--d, and if the two pairs of vertexes happen to be unconnected, to mix the pairs such that the they become connected as a--c $\&$ b--d. This switching was done $2L^2$ times for each random network.

\subsection{Social Networks}

For social networks, we have followed the algorithm described by \cite{Toivonen2006Model}. More precisely, we have started with a single seed vertex, and followed the below procedure $L-1$ times:

\begin{enumerate}
\item pick on average $m_r$ random vertices as initial contacts
\item pick on average $m_s$ neighbours of each initial contact as secondary contacts
\item connect a new vertex to the initial and secondary contacts
\end{enumerate} 

In the above, $m_r,m_s$ are random variables, re-evaluated for each new vertex, by using the expressions $\ceil{z+0.05+f_1\cdot v} $ and $\lceil n+0.05+f_2\cdot q\rfloor$. Here, $z,v,q$ are random variables uniformly distributed between 0 and 1, $n$ is uniformly distributed on the set $\{ 0,1,2,3\}$, and the $\lceil \cdot \rfloor$ is the rounding operator. Please note that these expressions reduce to those mentioned in \cite{Toivonen2006Model} for $f_1=f_2=0$.

\subsection{Alternative Update Rule}

To investigate whether the results in this paper, principally the structure dependence, depend crucially on the chosen update rule (\ref{eq:update}), we have conducted a smaller study in which (\ref{eq:update}) had an additional noise term:

\begin{equation}
x_i \rightarrow x_i(1-x_i)(f_C-f_D)dt + A\mu \ , \ \mu \sim N(0,1) \label{eq:update_noisy}
\end{equation}

\begin{figure}%
\centering
\includegraphics[width=.65\linewidth]{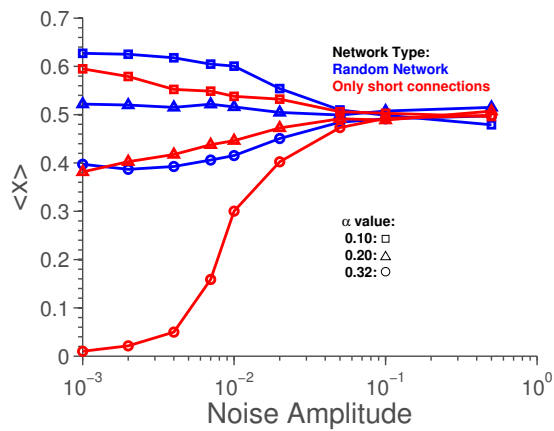} 
\caption{The effect on \meanx as a function of noise amplitude ($A$ in (\ref{eq:update_noisy})). We see that \meanx is different depending on the network structure, until the noise becomes so strong that the model collapses, as indicated by the disappearance of the $\alpha$-dependence.}
\label{fig:noisy}%
\end{figure}

This was chosen to mimic other update rules such as the Fermi update rule \cite{Szabo1998Evolutionary,Souza2009Evolution,Traulsen2007Pairwise}, in which suboptimal decisions are possible. By varying $A$, it is possible to investigate at what amount of noise is critical for our findings. In Fig \ref{fig:noisy} is seen \meanx as a function of noise amplitude ($A$) for two different network types and 3 different $\alpha$-values. We see that the structure-induced difference persists until complete model breakdown (when $\alpha$ ceases to be important, meaning that the relationship between cost and reward no longer has any influence). We also see that this breakdown occurs around $A \approx 0.01$, which seems very reasonable given that $dt=0.01$.

\bibliographystyle{unsrtnat}%

\end{document}